%% file: main.tex
\documentclass[journal]{IEEEtran}
\usepackage[english = american]{csquotes} % Makes "" work properly
\usepackage{url,multirow,comment,hyperref,amsmath,graphicx,amsmath,subfigure,listings,tabularx,psfrag,wrapfig,cite,verbatim,array, algorithm, algorithmicx, algpseudocode}
\usepackage[
locale = DE % comma as decimal mark
]{siunitx} %riad added this
\usepackage{caption} 
\usepackage{xcolor}
\begin{document}

%%%%%%%%%%%%%%%%%%%%%%%%%%%%%%%%%%%%%%%%%%%%%%%%%%%%%%%%%%%%%%%%%%%%%%%%%%%%%%%%
%%%%%%%%%%%%%%%%%%%              TITLE SECTION              %%%%%%%%%%%%%%%%%%%%
%%%%%%%%%%%%%%%%%%%%%%%%%%%%%%%%%%%%%%%%%%%%%%%%%%%%%%%%%%%%%%%%%%%%%%%%%%%%%%%%
%\title{LC Resonant Clock Resource Minimization using Compensation and sizing Decoupling Capacitance}
\title{Resonant Energy Recycling SRAM Architecture}

%\title{Low Power Cc-enhanced LC Resonant Clock Design Essentials}

%%%%%%%%%%%%%%%%%%%             AUTHORS SECTION             %%%%%%%%%%%%%%%%%%%%
\author{
%Anonymous for review purposes. Do not distribute.
Riadul~Islam,~\IEEEmembership{Member,~IEEE},
%Hany~A.~Fahmy,~\IEEEmembership{Student Member,~IEEE},
%Ping~Y.~Lin,~\IEEEmembership{Student Member,~IEEE},
Biprangshu~Saha,
~and Ignatius~Bezzam,~\IEEEmembership{Member,~IEEE}
%<-this stops a space

\thanks{R Islam is with the Department 
of Computer Science and Electrical Engineering, University of Maryland, Baltimore County, 
MD 21250, USA e-mail: {riaduli@umbc.edu}.}
\thanks{B Saha is with the Si2Chip Technologies, 
Road 1B, Gayatri Tech Park, Bengaluru, Karnataka 560066, India e-mail: {biprangshu.saha@si2chip.com}}
\thanks{I Bezzam is with the Rezonent Inc., 
1525 McCarthy Blvd, Milpitas, CA 95035, USA e-mail: {i@rezonent.us}}

%\thanks{This work was supported in part by the National Science Foundation
%under grant CCF-1053838.}
\thanks{This work was supported in part by Rezonent Inc. and by the UMBC startup grant.}
\thanks{Copyright (c) 2020 IEEE. Personal use of this material is permitted. 
However, permission to use this material for any other purposes must be 
obtained from the IEEE by sending an email to pubs-permissions@ieee.org.}
}
% The paper headers
\markboth{IEEE Transactions on Circuits and Systems--II}
%\markboth{IEEE Transactions on Computer-Aided Design of Integrated Circuits and Systems}%
{Shell \MakeLowercase{\textit{et al.}}: ??????}

\newcommand{\fixme}[1]{{\Large FIXME:} {\bf #1}}

% make the title area
\maketitle

%%%%%%%%%%%%%%%%%%%             ABSTRACT SECTION            %%%%%%%%%%%%%%%%%%%%
\begin{abstract}
Although we may be at the end of Moore's law, lowering chip power consumption is still the primary driving force for the designers.
To enable low-power operation, we propose a resonant energy recovery static random access memory (SRAM).
We propose the first series resonance scheme to reduce the dynamic power consumption of the SRAM operation. 
Besides, we identified the requirement of supply boosting of the write buffers for proper resonant operation.
We evaluated the resonant 144KB SRAM cache through SPICE and test chip using a commercial 28nm CMOS technology.
The experimental results show that the resonant SRAM can save up to 30\% dynamic power at 1GHz 
operating frequency compared to the state-of-the-art design.

\end{abstract}
\IEEEpeerreviewmaketitle

%\fixme{Pick a slightly different title than the paper to distinguish it.}
%%%%%%%%%%%%%%%%%%%           INTRODUCTION SECTION          %%%%%%%%%%%%%%%%%%%%
\input{introduction}

%%%%%%%%%%%%%%%%%%%            BACKGROUND SECTION           %%%%%%%%%%%%%%%%%%%%
\input{background}

%%%%%%%%%%%%%%%%%%%           PROPOSED DESIGN SECTION          %%%%%%%%%%%%%%%%%%%%
\input{propodesign}

%%%%%%%%%%%%%%%%%%%           PROPOSED DCM CDN SECTION          %%%%%%%%%%%%%%%%%%%%
%\input{proposed_dcm_cdn}

%%%%%%%%%%%%%%%%%%
%\input{proposed_dcmcs}

%%%%%%%%%%%%%%%%%%%           ANALYSIS SECTION          %%%%%%%%%%%%%%%%%%%%
\input{analysis}

%%%%%%%%%%%%%%%%%%%             RESULTS SECTION        %%%%%%%%%%%%%%%%%%%%%%%%%
%\input{noise_and_reliability}

%%%%%%%%%%%%%%%%%%%             Conclusions SECTION          %%%%%%%%%%%%%%%%%%%
\input{conclusion}

%%%%%%%%%%%%%%%%%%%             REFERENCES SECTION          %%%%%%%%%%%%%%%%%%%%

%\section*{Acknowledgments}
%The author would like to thank Professor Matthew Guthaus, H Fahmy, 
%and P Lin from UCSC for scientific discussion, software support, and inputs. 

\bibliographystyle{IEEEtran}
\bibliography{main}

\end{document}

%% file: introduction.tex
\section{Introduction}
\label{sec:intro}

%\textcolor{blue}{}
%Introduction:
Connecting an unlimited amount of high-speed embedded memories such as static
random access memories (SRAMs) to the microprocessor or a system-on-chip (SOC)
and having them as piggyback on computing is playing a pivotal role in
designing a high-performance computing system and data centers.  The embedded
memory consumes the significant portion of a microprocessor and enjoys more
aggressive design rules compared to the rest of the logic. % on a semiconductor chip. 
%The processor speed, which is much faster than the main memory (i.e.,
%dynamic-RAM or DRAM), consumes significantly more area and power per/bit than
%DRAM~\cite{Rusu:2003}. %As a result,\textcolor{blue}{[hussain-2]}. 
%next few lines briefed for space 
%While cache memories remain in the critical path of general-purpose computing and we
%observed a steady increase of embedded memories in modern microprocessors.
%Designing such a huge memory with a bounded performance and power budget
%becomes a very thorny problem that needs to be dealt with immediately and carefully.
However, the cache memories remain in the critical path of a general-purpose computing 
and designing large SRAMs with a bounded performance and power budget
becomes a very thorny problem that needs to be dealt with immediately and
carefully.

Among all the memories in a cache architecture, the SRAMs are essential for
efficient program execution. The SRAM provides the performance that is close to
the processor speed, which is much faster than the main memory; however, it
consumes significantly more area and power per/bit than dynamic-RAM or DRAM. 
%next few lines briefed for space 
%Due to this large
%size and high-speed, SRAMs consumes a significant amount of power in a
%microprocessor power-arc, which is considered as 10\%--20\% of total dynamic
%power~\cite{Rabaey:2009}. 
Due to large size and high-speed, SRAMs consumes about 10\%--20\% of total dynamic
power in a microprocessor power-arc~\cite{Rabaey:2009}. 
To reduce microprocessor power, researchers applied many low-power techniques;
among them, resonant
energy recovery (ER) clocking is used widely. % to reduce power. 
In this
work, we introduce resonant SRAM architecture to reduce effectively 
SRAM power even for non-cyclical operation and consequently 
enable ultra-low-power computing.
%talk little more about your contribution

\subsection{Prior Work and Motivations}
\label{subsec:prior_work}
%next few lines removed for space
%In general, the cache memory is realized by the SRAMs, primarily due to its
%compatibility with the standard CMOS logic and the high bandwidth. 
An SRAM consists of an array of data storage cells and peripheral circuits to control
the memory and allow us to read/write with a bit-level precision.  
The SRAM's reliability depends on the cell's robustness and the
peripheral circuitry to noise; and process, supply voltage, and chip
temperature (PVT) variation. Besides, researchers identified that a
significant amount of dynamic and static power consumed by SRAMs, especially at
sub 10nm technology node with increased SRAM density and many SRAM cuts in a
single chip.  As a result, there has been a tremendous amount of work on SRAM
design to improve SRAM design
efficiency~\cite{Karl:2016,Maroof:2017,Joshi_cicc:2017}.  %~\cite{Wei:1992,Karl:2016,Joshi:2017,Maroof:2017,Joshi_cicc:2017,Koo:2015}.
However, this work's primary goal is to reduce the SRAM power without affecting
the cell density and performance of the memory.

The most widespread low-power techniques in IC design are dynamic voltage and
frequency scaling (DVFS)~\cite{Nowka:2002}, resonant LC
clocking~\cite{Bezzam:2015,Fuketa:2014,Sathe:2013}, current-mode (CM)
clocking~\cite{Islam:2015}, and etc. Among different low-power techniques, LC
resonant clocking is very interesting due to its constant phase and magnitude.
However, in the proposed research, we apply LC resonance to reduce SRAM power
consumption.
%resonant clock The resonant energy recovery clocking has great potential to
%reduce CDN power.  
Previously, researchers applied resonant clocking in SRAMs to save
power~\cite{Tzartzanis:1996}. This method used resonant ER
latches in the address, wordline, and input latches to save energy.

%This method applied resonant energy recovery
%latches in the address latches, wordline latches, and input latches to save
%significant amount of dynamic power.

\begin{comment}
Other researchers proposed resonant energy recovery clocking to reduce CDN power.                                         
Resonant clocking uses the clock capacitance and an on-chip inductor to resonate at a central frequency. The basic idea of resonant clocking is that it saves power by storing and recycling energy into the magnetic field of an inductor and the electric field of a capacitor~\cite{Esmaeili:2012,Rahman:2018}. However, resonant clocking suffers from a large slew rate due to the sinusoidal nature of the produced signal and consumes high short-circuit power. In addition, it requires a high quality factor for the on-chip inductor, which is hard to maintain over a large frequency band. Recently, intermittent-resonant clocking solved both of the issues by using extra clock driver circuitry~\cite{Rahman:2018}. However, all of these resonant clocking schemes require passive devices and additional bias circuits.
Another interesting idea reduces resonant clock passive device resources, perhaps very relevant to our work. However, it suffers from frequency scaling issues~\cite{Lin:2015}.
\end{comment}
%CM clock     
               
\begin{figure}[t!]
	\includegraphics[width = 0.45\textwidth]{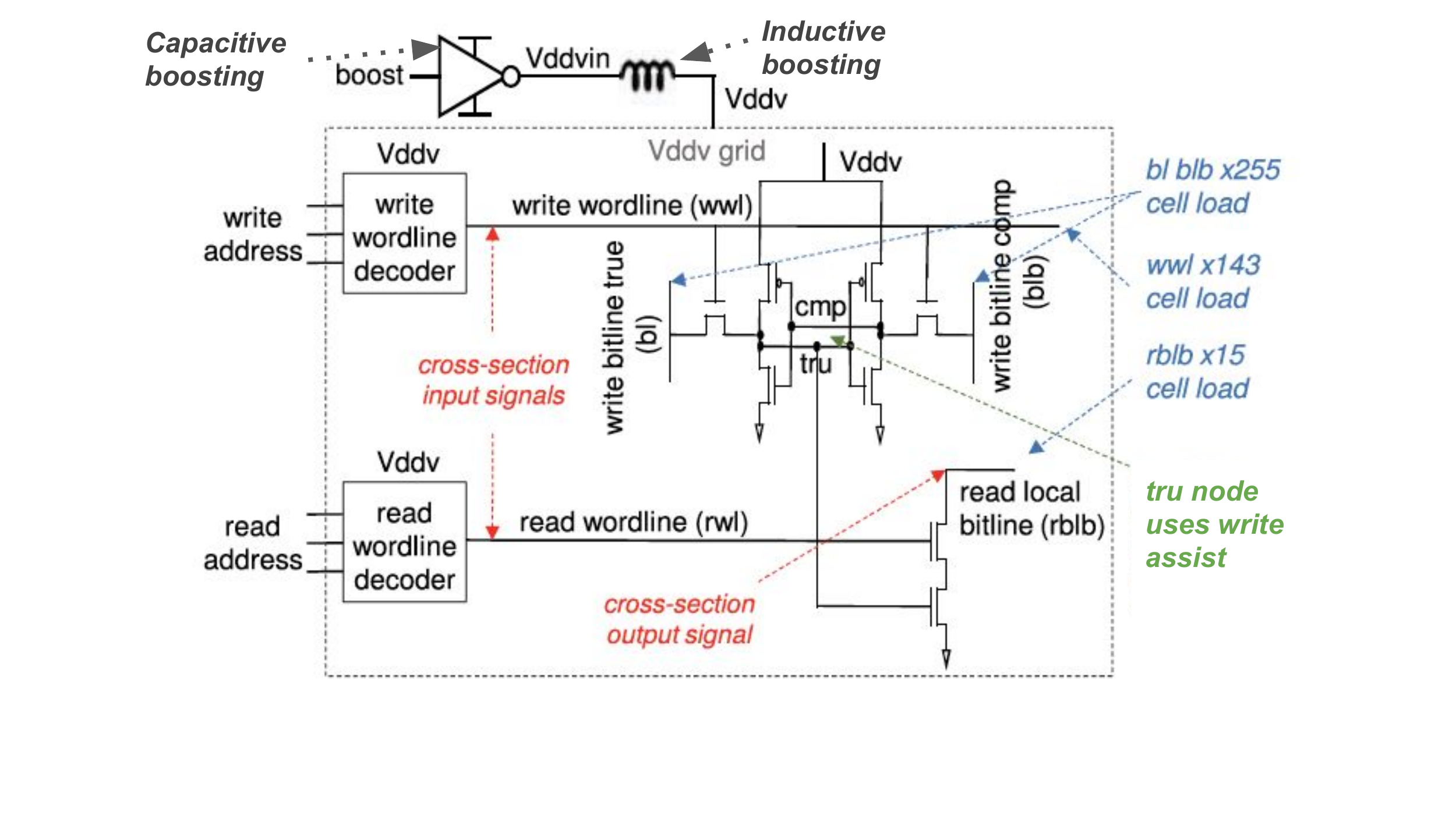}
	\captionsetup{aboveskip=-0.10cm,belowskip=-.25cm}
	\caption {To enable low supply voltage operation researchers applied 
both capacitive and inductive boosting of the input voltage; figure modified from~\cite{Joshi:2017}.}
	\label{fig:ibm_supply_boost}
	\vspace{-0.35cm}
\end{figure}

One of the recent works, researchers applied supply boosting for SRAMs as a
combination of capacitive and inductive boosting, as shown in
Figure~\ref{fig:ibm_supply_boost}~\cite{Joshi:2017}. This approach uses a
transistor-based capacitive coupling for initial supply boosting on 14-nm SOI
FinFET technology.%~\cite{Joshi:2017}. %reference removed for space
The initial enhanced supply voltage
further amplified using a resonant inductor to achieve a meager 0.3V supply
voltage operation. This method uses two additional transistors per cell as
reading and write-assist, compared to the standard six transistors (6T) SRAM
cell. However, this method requires a sizeable 4nH inductor for $144 \times 256$b 
SRAM for 0.3V operation. Another similar approach uses
%SRAM block for 0.3V operation. Another similar approach uses
novel cascaded inductive and capacitive booster to reduce 8T SRAM supply
voltage down to 0.24V considering 14nm SOI FinFET
technology~\cite{Joshi_cicc:2018}. However, there is no real guideline on
the appropriate inductor size for the 25.5Kb SRAM.

To enable low-power memory operation, we introduce resonant bit lines and
inductive supply boosting for the write drivers using conventional 6T SRAM
cells on a 28nm CMOS technology.  We applied the proposed technique on a
generic K2 cryptoprocessor (GCrP) with 144KB of SRAM, which enables up to 20\%
lower SRAM dynamic power compared to the conventional CMOS SRAM implementation
with no leakage penalty.

\subsection{Main Contributions}
\label{subsec:main_contributions}
In this work, we reduced the embedded SRAM power by introducing series
resonance on the cache memory. In particular, the critical contributions of
this work are:
\begin{itemize} \renewcommand{\labelitemi}{$\bullet$}
	\item The first series resonant SRAM architecture. 
	\item The first inductor sizing technique considering discharge time and maximum resonant swing.
	\item The significant reduction of the SRAM dynamic power without changing the conventional 6T cell architecture.
\end{itemize}

\subsection{Paper Organization}
\label{subsec:paper_org}
The rest of this paper is organized as follows. In
Section~\ref{sec:resonant_background}, we first introduce the resonant
techniques.  Section~\ref{sec:proposed_sram} presents the proposed SRAM
architecture. In Section~\ref{sec:analysis}, the power efficiency of the
proposed resonant design with existing industry-standard schemes is
investigated with simulation and experimental results. Finally,
Section~\ref{sec:conclusion} concludes the paper.

%% file: background.tex
\section{Resonant Background}
\label{sec:resonant_background}

%removed for space
%Resonant clocking is one of the most popular low-power techniques.
The ER resonant clocking can be classified as standing wave~\cite{O'Mahony:2003}, 
rotary~\cite{Taskin:2006}, and LC resonant~\cite{Islam_res:2018, Bezzam:2015}. 
Among various resonant schemes,
rotary clocks have fixed amplitude but a variable phase. In contrast, a
standing wave clock has a constant phase but a varying amplitude. LC resonants
mimic the conventional CMOS clocking accurately with a higher slew rate;
however, it exhibits tremendous potential to save dynamic power.

\begin{figure}[t!]
\vspace{-0.35cm}
	\includegraphics[width = 0.34\textwidth]{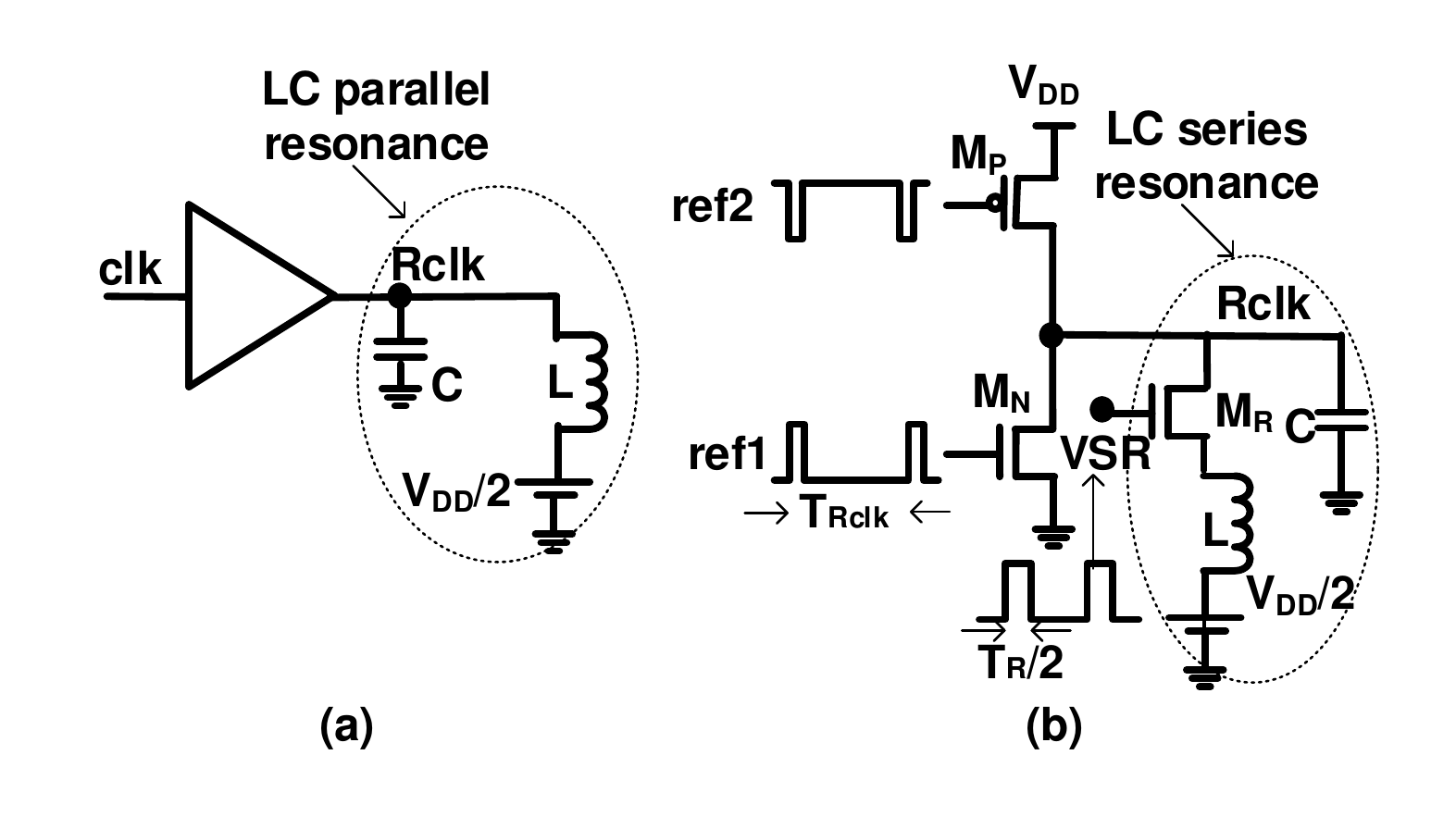}
	\captionsetup{aboveskip=-0.10cm,belowskip=-.25cm}
	\caption {(a) Parallel resonance exhibits power saving at a limited frequency range~\cite{Sathe:2013}, (b) A series resonance uses pulsed signals to maintain rail-to-rail voltage swing, and the switched controlled inductor helps it to operate efficiently in a wide frequency range~\cite{Bezzam:2015}.} 
	\label{fig:resonant}
	\vspace{-0.35cm}
\end{figure}

In a conventional CMOS design, half of the switching energy is wasted in
charging a capacitive node (i.e., 0-to-1 transition); and the other half is
wasted in the discharging phase (i.e., 1-to-0 transition). The LC resonance
stores some of the discharge energy in the magnetic field on an inductor (L)
and recycles during the charging phase to charge the capacitor (C). To maintain
the resonance, we need an external source to compensate for the resistive loss.
%removed for space
%and the LC circuit's inductive and capacitive reactances cancel out each other.
LC resonant clocking can be categorized as parallel and series resonance. At
resonance, conventional LC parallel resonance cancels out inductive and
capacitive reactance, as shown in Figure~\ref{fig:resonant}(a).

On the other hand, the series LC resonance requires additional transistor
switching ($M_R$), as shown in Figure~\ref{fig:resonant}(b). The pull-up
($M_P$) and pull-down ($M_N$) switches help maintain rail-to-rail voltage
operation. The ($VSR$) signal is generated from the rising and falling edges
of the input clock (clk) signal. The primary advantage of series resonance
compared to the parallel resonance is the wideband frequency operation. 
The required timing of $T_R$ is only a fraction of the overall 
resonance clock ($T_{Rclk}$), and inductor sizes are an order of magnitude less.

%commented for clarifications
%The primary advantage of series resonance
%compared to the parallel resonance is the wideband frequency operation due to
%the required timing of $T_R$ is only a fraction of the overall resonance
%clock ($T_{Rclk}$) and inductor sizes are an order of magnitude less.

%caption curtailed for space
%(a) Parallel resonance uses a buffer to drive the network and exhibits power saving at limited frequency range~\cite{Sathe:2013}, (b) A series resonance uses pulsed signals to maintain rail-to-rail voltage swing and the switched controlled inductor helps it to operate efficiently in wide frequency range~\cite{Bezzam:2015}.

\begin{figure}[t!]
	\includegraphics[width = 0.40\textwidth]{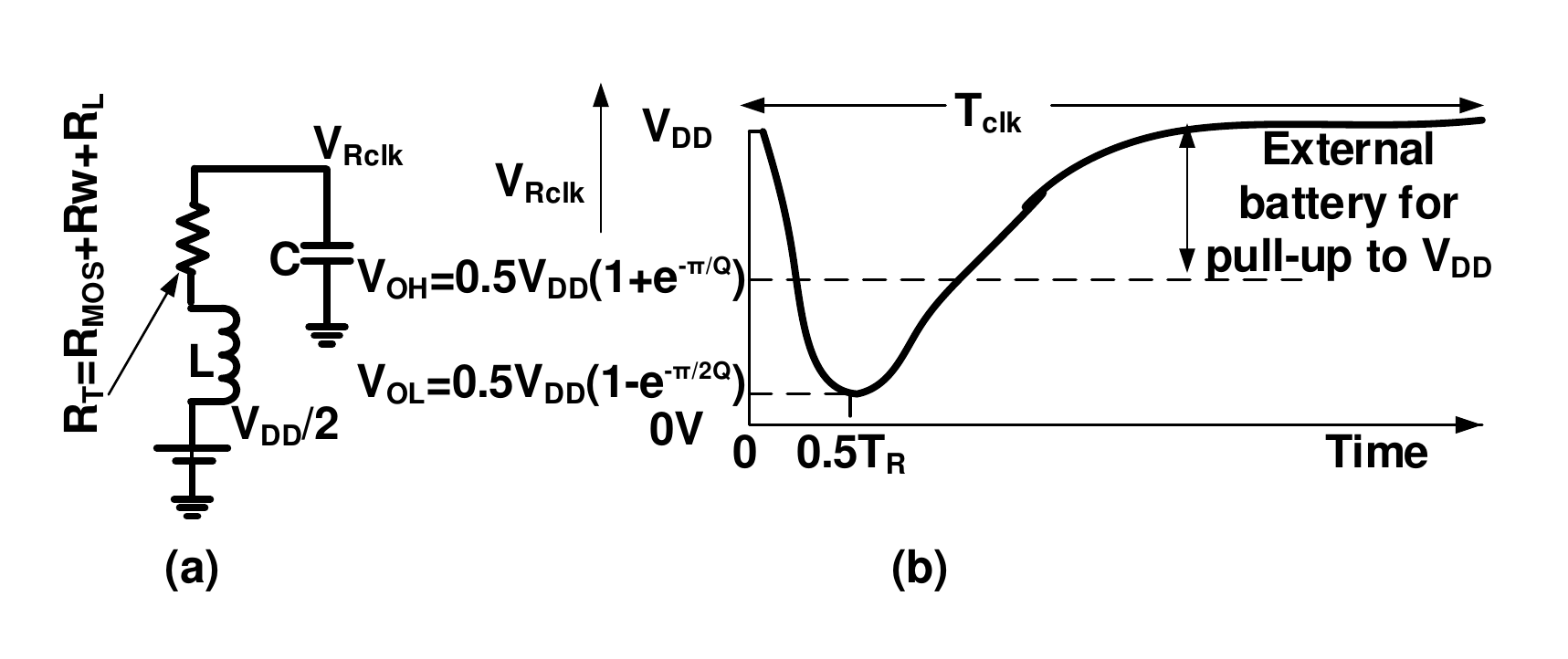}
	\captionsetup{aboveskip=-0.10cm,belowskip=-.25cm}
	\caption {(a) The series resonance equivalent circuit model help us to identify the proper $R_TLC$, (b) The output capacitive voltage is important to identify maximum resonant $V_{Rsw}$ and timing specification of a design.} 
	\label{fig:rlc}
	\vspace{-0.35cm}
\end{figure}

%removed for space
%A series resonance equivalent circuit helps us accurately define the resonant frequency and the corresponding inductor size. Figure~\ref{fig:rlc}(a) shows the equivalent series RLC network for the Figure~\ref{fig:resonant}(b). 
Figure~\ref{fig:rlc}(a) shows a series resonance equivalent circuit which 
helps us accurately define the resonant frequency and the corresponding inductor size. 
The equivalent total resistance ($R_T$) is the combination of  ``ON'' NMOS 
resistance ($R_{MOS}$), wire resistance  ($R_W$), and inductor parasitic 
resistance ($R_L$). According to Kirchhoff's voltage law (KVL), we can write,
\begin{equation}
R_Ti_L(t) + \int \frac{i_L(t)}{C}dt + L \frac{di_L(t)}{dt} =  \frac{V_{DD}}{2}  
\label{eq:kvl}
\end{equation}
where $i_L$ is the inductor current~\cite{Bezzam:2015}. %For zero initial inductive current, 
%removed for space
%For zero initial inductive current, 
%we can express a second order differential equation for the $i_L$ as,
%\begin{equation}
%\frac{d^2i_L}{dt^2} + \frac{R_T}{L} \frac{di_L}{dt} + \frac{i_L}{LC} = 0  
%\label{eq:diff_iL}
%\end{equation}
The minimum inductance required considering underdamped condition is $L>\frac{R^2_TC_L}{4}$. 
This is a critical condition that helps us to pick the right inductor for our design. 
From the Equation~\ref{eq:kvl}, we can express the $i_L$ as, %for zero initial inductive current as,
\begin{equation}
i_L(t) = \frac{V_{DD}}{2\sqrt{\frac{L}{C}}\sqrt{1-\frac{1}{4Q_f^2}}}e^{-\frac{tR_T}{2L}}sin(2\pi f_Rt)  
\label{eq:iL}
\end{equation}
where $f_R=\frac{1}{T_R}=\frac{1}{2\pi} \sqrt{\frac{1}{LC}-\frac{R_T^2}{4L^2}}$ represent the damping oscillation frequency, 
$Q_f=\frac{1}{R}\sqrt{\frac{L}{C}}$ is the quality factor.
%removed for space
%$Q_f=\frac{1}{R}\sqrt{\frac{L}{C}}$ is the quality factor, and can be represented as,
%\begin{equation}
%f_R = \frac{1}{2\pi} \sqrt{\frac{1}{LC}-\frac{R_T^2}{4L^2}}  
%\label{eq:fR}
%\end{equation}
%The Equation~\ref{eq:fR} helps us to identify the proper inductor, $R_T$, and capacitive 
%load for corresponding damping frequency in Section~\ref{sec:proposed_sram}. 
The $f_R$ value help us to identify the proper inductor, $R_T$, and capacitive 
load for corresponding damping frequency in Section~\ref{sec:proposed_sram}. 
The $T_R$ time corresponds to the bitline discharge time and will be discussed in detail in 
Section~\ref{sec:analysis}. Now, we can compute the voltage across the capacitor ($V_{Rclk}$) as,
\begin{equation}
%\begin{split}
\begin{aligned}
V_{Rclk}(t) = {} &\frac{V_{DD}}{2} + \frac{V_{DD}}{2} e^{-\frac{tR_T}{2L}}cos(2\pi f_Rt) \\ 
& - \frac{1}{2Q_f}  \frac{V_{DD}}{2} e^{-\frac{tR_T}{2L}}cos(2\pi f_Rt)
\label{eq:VRclk}
%\end{split}
\end{aligned}
\end{equation}
Figure~\ref{fig:rlc}(b) shows the $V_{Rclk}$ curve, where the difference between resonant high output ($V_{OH}$) and low output ($V_{OL}$) represent the voltage-swing ($V_{Rsw}$) due to the autonomous ER. Hence, we performed extensive simulations to identify the proper inductor and capacitive load to maximize this $V_{Rsw}$ in Section~\ref{subsec:sim}. 

%removed for space
%(a) The series resonance equivalent circuit model help us to identify the proper inductor size, total resistance, and capacitive load, (b) The output capacitive voltage is important to identify maximum resonant $V_{Rsw}$ and timing specification of a design.}

%% file: propodesign.tex
\section{Proposed Resonant SRAM Architecture}
\label{sec:proposed_sram}
	 	 	
To improve the power-performance of embedded cache memory, we propose the
resonant SRAM architecture. Empirically, in an SRAM write operation, many bit
cells switches and make it the most dynamic power-consuming phase as all the bitlines with
high capacitive load discharges during the write operation and charges back,
irrespective of mux factor. Unlike conventional low supply voltage SRAMs, we
recycle the discharged energy from the bitline load capacitances.  The proposed
architecture uses an on-chip inductor is conditionally attached to bitlines
biased by another supply voltage ($\frac{V_{DD}}{2}$) to store the discharge
energy considering in series resonance topology. While charging back in the
recovery phase, the stored energy is used to charge the load capacitance
towards $V_{DD}$. The inductor in the series resonance circuit stores the
discharge energy in the form of a magnetic field, which in turn empties the
load capacitance charge to a greater extent, hence stores the electric charge
in $\frac{V_{DD}}{2}$ node. In the recovery phase, the same series resonance
circuit pulls out the same amount of charge from the $\frac{V_{DD}}{2}$ node,
leaving zero net currents from that node in the whole cycle, ensuring no
additional power drawn from the inductor bias supply.

\begin{figure}[t!]
	\includegraphics[width = 0.4\textwidth]{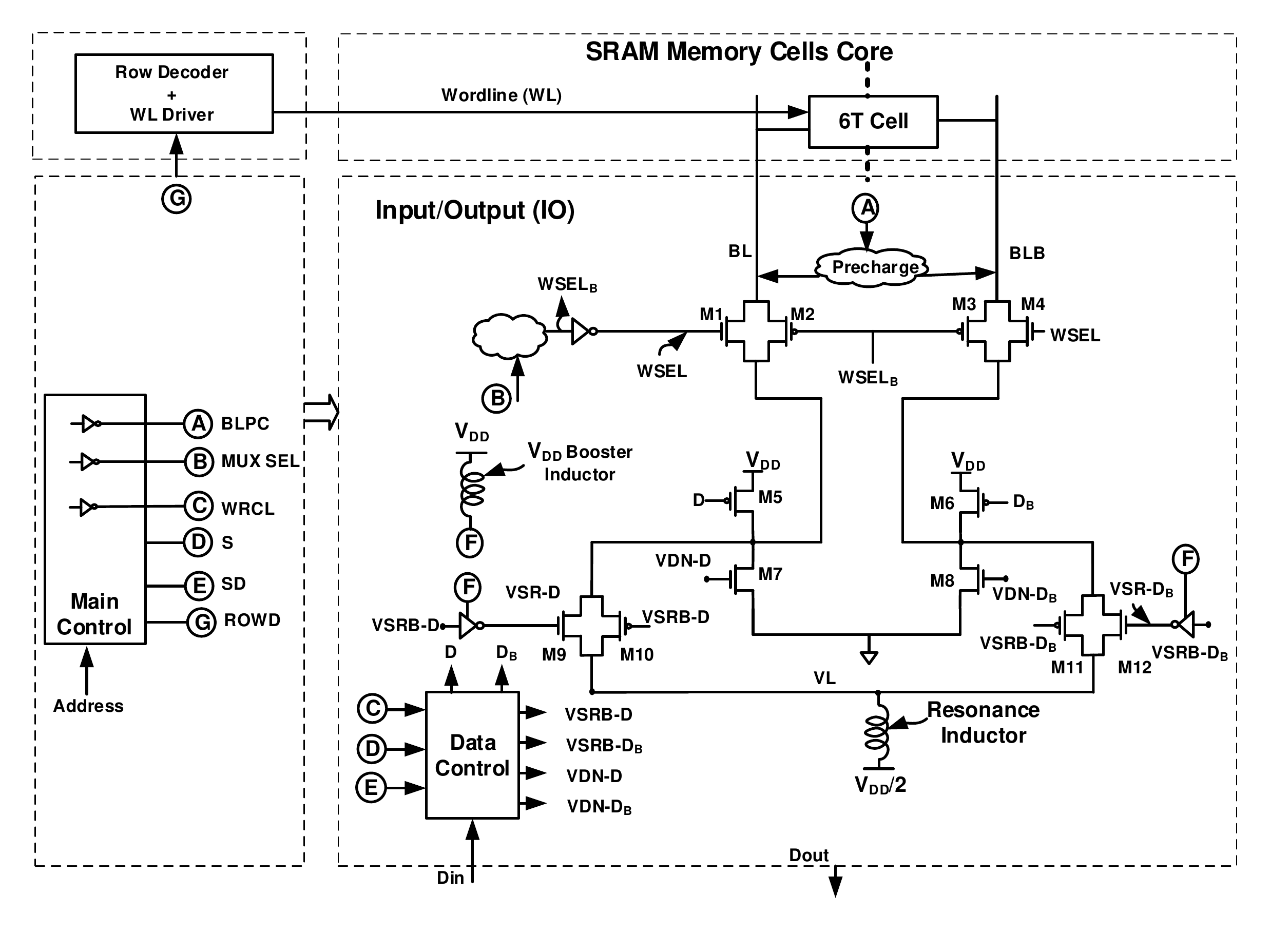}
	\captionsetup{aboveskip=-0.00cm,belowskip=-.0cm}
	\caption {Unlike existing low-power SRAMs~\cite{Joshi:2017,Joshi_cicc:2018}, we proposed the first series resonant 
SRAM architecture that uses ER bitlines and inductive 
boosting for the write driver to sustain resonance at a wide frequency range.} 
	\label{fig:proposed_sram}
	\vspace{-0.25cm}
\end{figure}

%Unlike existing low-power SRAMs~\cite{Joshi:2017,Joshi_cicc:2018}, we proposed the first resonant 
%SRAM architecture that uses energy recovery resonant bitlines and inductive 
%boosting for the write driver supply voltage to sustain resonance at a wide frequency range efficiently.

%removed for space
%The proposed SRAM architecture is shown in Figure~\ref{fig:proposed_sram}. The
%proposed architecture is based on the existing 6T storage cell and uses
%conventional read-write methodology with peripheral circuitry. 	 
%The proposed SRAM architecture is shown in Figure~\ref{fig:proposed_sram}. 
The proposed SRAM architecture is based on the existing 6T bit cell and uses
conventional read-write methodology with peripheral circuitry, as shown in Figure~\ref{fig:proposed_sram}. 	 	 	
%Write driver architecture
The write driver connects to bitlines load through the transmission gates
(M9-M12) and controlled by VSR-D and VSRB-D and their complements. To enable
series resonance inductor is placed between node VL and write driver
transmission gates. We ensure the full rail-to-rail swing using a pair of NMOS
transistors (M7-M8) parallel to the write driver's series resonance
transmission gates. To achieve a reasonably high $Q_f$, we use
low threshold voltage devices in the series resonance path (M1, M9, M4, and
M12).   
	 	 	
%Shared inductor architecture- 
It is vital to use a shared inductor to reduce the size of the inductor. We
need N number of write drivers for N number of bits. A shared inductor connects
all the write drivers to the VL node. Hence total load capacitance increases N
folds, and series path effective resistance (M1-M2 and M9-M10) decreases N
folds. The proposed architecture makes it possible to achieve the target
frequency of operation with a low value of the inductor and high $Q_f$.  
	 	 	
%Extra Overdrive Voltage for the series Path transistors (Nmos)- 
As the transistor source is connected to the $\frac{V_{DD}}{2}$ node through
the inductor, driving the gate of these NMOS transistors %(i.e., transmission gates) 
by $V_{DD}$ turns these ``ON'' loosely. To overcome this problem, we
generate a bump voltage with another resonant path to get the bump 
without wasting the power. The M9 and M12 transistor gate capacitances
are the load capacitance for this path. For all the write drivers, we used a
shared booster inductor.
%removed for space
%The NMOS transistors (M9 and M12) gate capacitances
%are the load capacitance for this path. For all the write drivers, we used a
%shared booster inductor.

%% file: analysis.tex
\section{Simulations and Test Chip Results}
\label{sec:analysis}

\subsection{Simulations Results}
\label{subsec:sim}
\begin{figure}[t!]
\vspace{-0.25cm}
	\includegraphics[width = 0.3\textwidth]{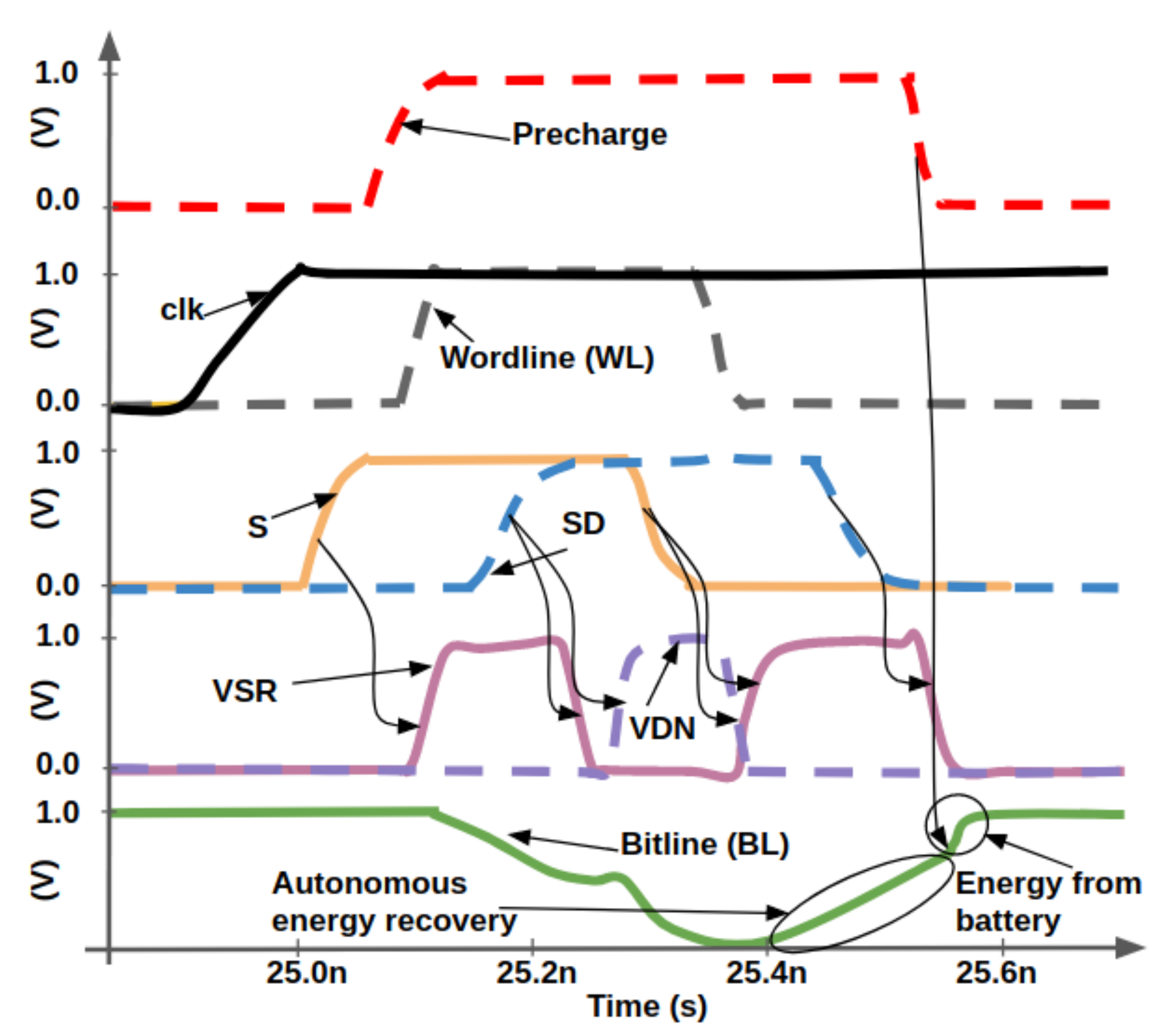}
	\captionsetup{aboveskip=-0.00cm,belowskip=-.0cm}
	\centering
	\caption {The internal control signals S and SD produce the VSR and VDN 
signals; the former helps the bitline to discharge on the resonant inductor path, 
while the latter confirms the full rail-to-rail swing of the bitline in the resonant write cycle.} 
	\label{fig:resonant_write} 
	\vspace{-0.35cm}
\end{figure}

%control circuitry
We performed extensive simulations of peripheral and SRAM core arrays using
TSMC 28nm CMOS technology. For series resonant operation, we generate two
voltage pulses (VSR and VSD) using the signal S and a delayed version of S,
named SD. Both the signals are generated from the SRAM input clk
signal. We generate the VSR signal using a 2-input XOR gate where the input
signals are S and SD. Similarly, we extract the VSD signal using a 2-input AND
gate where the input signals are S and SD.  During the write operation, VSR and
its complementary signals discharge the bitlines using transmission gate M9-M12
transistors through the inductive path, as shown in
Figure~\ref{fig:proposed_sram}. However, to fully discharge the bitline, we
need the VDN signal. Figure~\ref{fig:resonant_write} shows the simulation
results of $512 \times 128$ bits resonant SRAM control signals during the write
operation. 

We computed the required $T_R$ pulse width depending on the SRAM MUX factor or
number of connected columns. According to our analysis, the $\frac{T_R}{2}$
time reduces with the reduction of the number of associated columns for a
fixed inductor. The Table~\ref{tab:Tr_time} shows the results of this analysis.
We can adjust the pulse width by controlling the delay between S and SD
signals. However, we may need to change the pulse width for proper resonant operation 
due to process variation. To tackle this issue, we have a 4-bit register 
controlled input signal to the tuned circuit that generates the S and SD signals.

%Booster Inductor Sizing??

%\begin{comment}

\begin{table} [t!] \large %\footnotesize
	\renewcommand{\arraystretch}{1.2}
	\caption{For a fixed resonant inductor, the $\frac{T_R}{2}$ time reduces with the reduction of the total number of associated columns.}
	\label{tab:Tr_time}
	\centering
	%\resizebox{9cm}{!}
	\scalebox{0.65}{
		{\begin{tabular}{|c|c|c|c|c|}
				\hline
				MUX factor    & \# of columns        & Total cap (pF)   &  Inductor (nH)  & $\frac{T_R}{2}$ (ps) \\
				\hline
				1             & 256                  & 10.10            & 0.621           & 248.0\\
				\hline
			        2             & 126                  & 5.07             & 0.621   	  & 176.0\\
			       \hline
			        4             & 64                   & 2.53             & 0.621   	  & 125.0\\
			       \hline

	\end{tabular}}}
	\vspace{-0.35cm}
\end{table}

%\end{comment}

\begin{figure}[t!]
	\includegraphics[width = 0.4\textwidth]{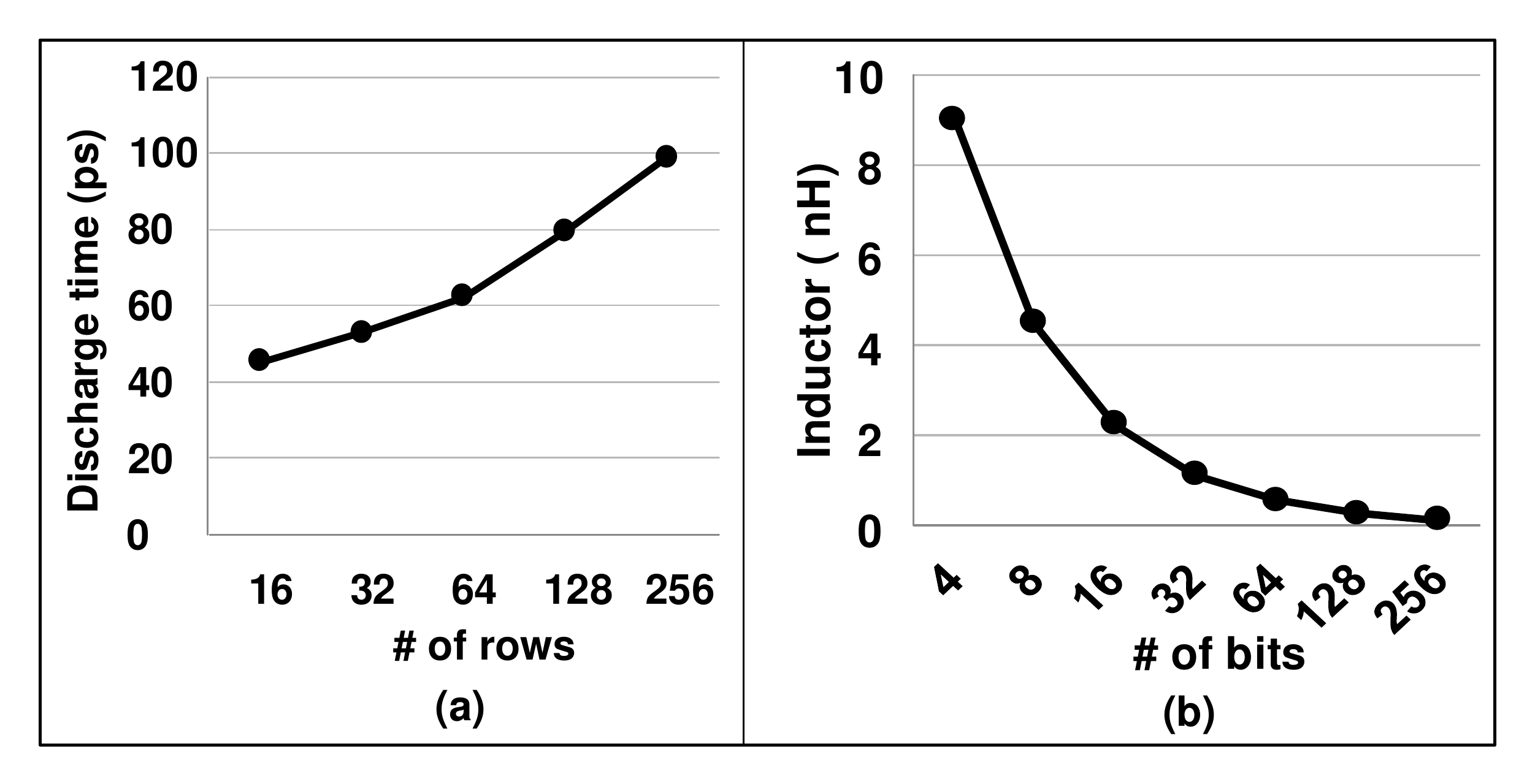}
	\captionsetup{aboveskip=-0.00cm,belowskip=-.0cm}
	\centering
	\caption {(a) The discharge time increases with the increase in the number 
of rows, and this analysis helps us to precisely define the $\frac{T_R}{2}$ time for 
operating a specific frequency, (b) we identified the resonant inductor sizing by the
varying number of bits for a target voltage swing and $\frac{T_R}{2}$ time.} 
	\label{fig:resonant_discharge} 
	\vspace{-0.45cm}
\end{figure}

The performance of resonant SRAM depends on the bitline discharge time. We
tuned the SRAM instance for a particular number of bits and varied the number
of rows to compute the discharge time. Figure~\ref{fig:resonant_discharge}(a)
shows the results of this analysis. As expected, the discharge time increases
with the increase of the number of rows. This analysis helps us to define the
$\frac{T_R}{2}$ time and optimize our design for a target frequency range.

%Resonant Inductor Sizing??
To properly size the resonant inductor, we consider a target frequency range of 200MHz-1GHz
and a bitline discharge time down to 100ps. Also, we use the target resonant
voltage swing is approximately two-third of the $V_{DD}$. We varied the number
of bits connected to the VL line of Figure~\ref{fig:proposed_sram} and
identified the inductor size that achieves the target design parameters.
Figure~\ref{fig:resonant_discharge}(b) shows the results of this analysis.
Clearly, increasing the number of bits reduces the inductance requirement. 
This analysis also helps us to achieve the target power gain in the form of
$V_{Rsw}$ even with the varying inductor size. The primary reason for
this constant voltage swing is the parallel write drivers' resistance reduces
with the reduction of inductance, resulting in a fixed $Q_f$.

%\textcolor{blue}{
We performed extensive simulations considering corner cases, supply voltage, and temperature variations. Using fast-fast (FF) devices, 1.1V supply voltage, the maximum bitcell write time is 37ps at $125^{\circ}C$., using slow-slow (SS) devices, 0.81V supply voltage, and the maximum bitcell write time is 79.6ps at $-40^{\circ}C$. %}

\subsection{Test Chip Experimental Results}
\label{subsec:exp}
\begin{figure}[t!]
%\begin{center}
%\centering
	%\includegraphics[width = 0.22\textwidth]{./figures/chip_die_gray.png}
	\includegraphics[width = 0.40\textwidth]{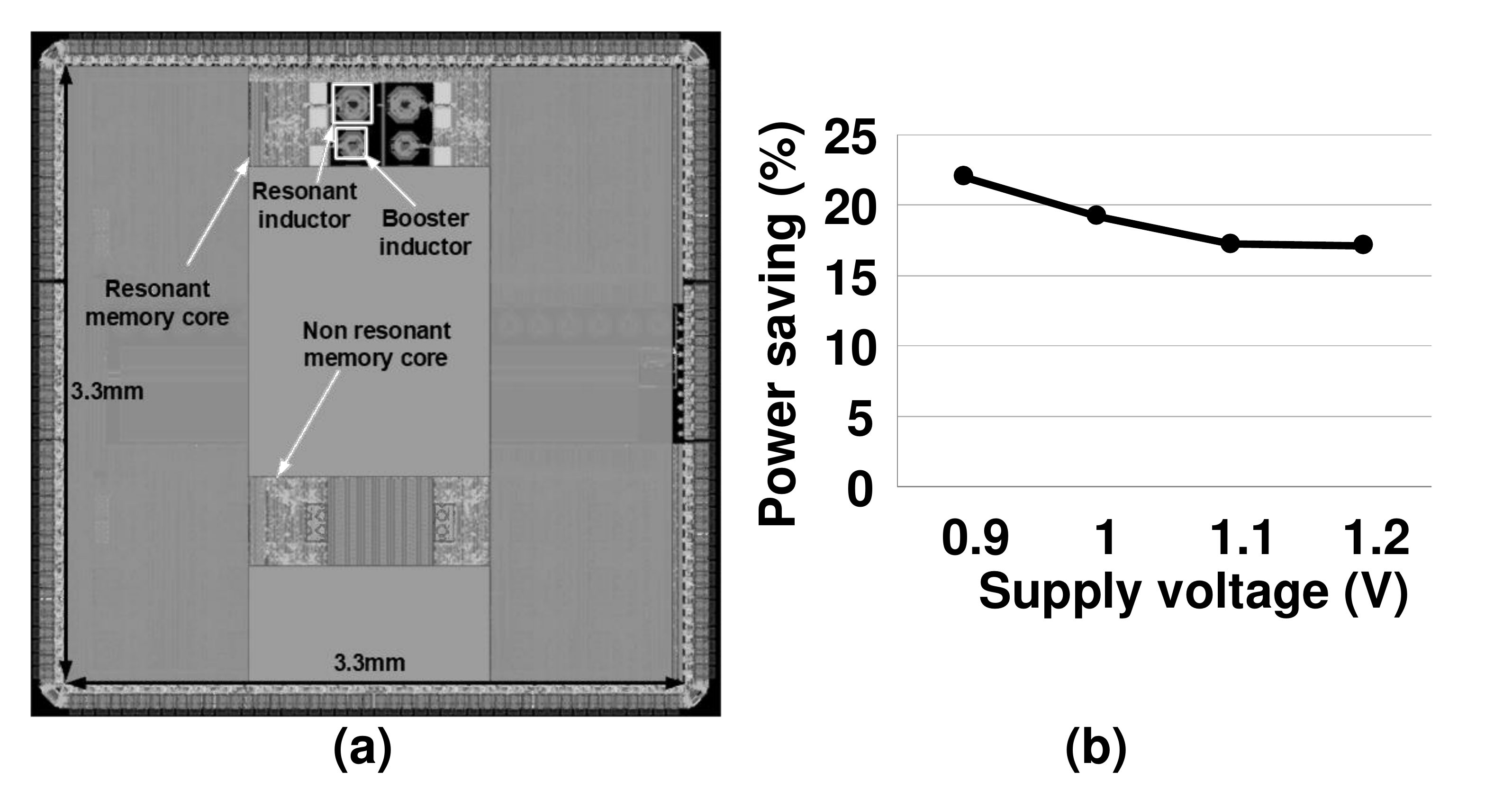} 
	\captionsetup{aboveskip=-0.00cm,belowskip=-.0cm}
	\centering
	\caption {(a) Die photograph of the test chip, (b) The proposed memory power saving decrease with the increase of supply voltage.}
	\label{fig:die_chip} 
	\vspace{-0.40cm}
%\end{center}
\end{figure}

We verified the proposed resonant SRAM architecture by designing a 
GCrP using a commercial 28nm CMOS technology. We integrate
144KB of SRAM for the GCrP, as shown in Figure~\ref{fig:die_chip}(a). For
comparison, we created a similar GCrP with the same amount of conventional
SRAMs. We embed a 2nH resonant inductor and 0.5nH booster inductor for each 8KB
of memory instance. 

The total resonant memory area is $0.36mm^2$, which consumes only 2\% extra
silicon area for the additional switching transistor 
that recycles the energy than the conventional 6T-based SRAM design. 
We used the top two metals for inductors. The primary
goal of this test chip is to verify the power efficiency of the resonant SRAM.
The resonant SRAM operating supply voltage ranges from 0.9V to 1.2V, which results in 
22\% to 17\% overall memory power saving compared to the conventional industry 
standard SRAM architecture with no leakage penalty, as shown in Figure~\ref{fig:die_chip}(b). The primary reason is the use of 
the same 6T cells. We set the chip operating frequency 200MHz to 1GHz, which results in 20\% 
to 30\% overall memory power saving compared to the non-resonant memory.

%% file: conclusion.tex
\section{Conclusion}
\label{sec:conclusion}

In this paper, we presented the first series resonant SRAM architecture to reduce
memory power. The proposed architecture uses a booster inductor for the write
drivers and a resonant inductor to recycle energy from the SRAM bitlines. We
fabricated a test chip using TSMC 28nm CMOS technology.  The proposed
resonant SRAM can save up to 30\% dynamic power with only a 2\% area penalty
than the conventional CMOS SRAM.s